# Coherence properties of infrared thermal emission from heated metallic nanowires

Levente J. Klein and Hendrik F. Hamann
IBM T.J. Watson Center, Yorktown Heigths, NY
Yat-Yin Au and Snorri Ingvarsson
University of Iceland, Iceland

Coherence properties of the infrared thermal radiation from individual heated nanowires are investigated as function of nanowire dimensions. Interfering the thermally induced radiation from a heated nanowire with its image in a nearby moveable mirror, well-defined fringes are observed. From the fringe visibility, the coherence length of the thermal emission radiation from the narrowest nanowires was estimated to be at least 20 µm which is much larger than expected from a classical blackbody radiator. A significant increase in coherence and emission efficiency is observed for smaller nanowires.

Blackbody radiation emitted by heated objects is generally considered incoherent and unpolarized with a broad spectrum described by Planck's law.[1] However, surface phonon polaritons can couple to the electromagnetic modes in polar materials and increases the coherence of thermal radiation.[2-4] Coherent thermal light sources are important in the structural characterizations of nanoscale molecular and biological samples[5] and development of efficient thermoelectric radiation sources.[6]

Recently it was shown that thermal radiation from individual heated metal nanowires can be highly polarized with antenna-like angular radiation pattern.[7] A significant radiation enhancement was observed for very narrow nanowires, which can be attributed to the increased spatial confinement of charge fluctuations along the long axis of the nanowires thereby reducing spatial decoherence effects.[7]

Here we investigate the coherence properties of the thermal emission from individual heated nanowires. The thermal radiation from fixed aspect ratio, heated metallic nanowires is self-interfered by creating an image in a moveable mirror in close proximity of the nanowire. The results are threefold; First of all, we observe distinct interference fringes at a mirror distance of more than 10 µm from the nanowire suggesting much longer coherence lengths than what is typically anticipated for thermal radiation from classical blackbodies. Second, the fringe visibility increases drastically with decreasing nanowire size and lower temperatures. Finally, and in agreement with earlier results, with an increasing fringe visibility a significant increase in emission intensity (thermal radiation signal per unit area) is observed.

Specifically, metallic Ti/Pt nanowires (thicknesses of 5 nm and 45 nm respectively) with a constant aspect ratio (i.e., length / width = 10) have been fabricated using electron beam lithography. The width of the nanowires ranges from 125 nm to 2 µm with a corresponding length between 1.25 and 20 µm, respectively. The nanowires are patterned on a 100 nm thick thermally grown $SiO_2$ layer on a Si substrate. The temperature dependent resistance of the nanowires allows precise control of temperature using a previously described measurement/control setup.[7] As illustrated in Fig. 1(a), individual nanowires are resistively heated to temperatures up to 600 °C. Their thermal radiation is projected from the transparent backside of the Si-wafer substrate onto an infrared camera (InSb detector sensitive in the 3-5 µm spectral region) using a reflective objective (35 x magnification and NA ~0.5, with a depth of focus of ~10 µm).

In order to interfere the thermal radiation, an Al coated mirror attached to a piezo tube is positioned in the close proximity of the nanowire.[8] The thermal radiation signal from the nanowire is then recorded as a function of mirror displacement ($\Delta z$) until a "contact" point. The optical path difference between the radiation from the nanowire and its image in the mirror is $2\Delta z$. Due to the finite depth of focus of our detection system only the radiation from the nanowire (but not from its image) is detected when the mirror is far away from the sample. As the mirror approaches the nanowire, the radiation from both the nanowire and its image contributes to the total detected signal. Thus, both an increased radiation intensity and a periodic modulation of intensity is observed.

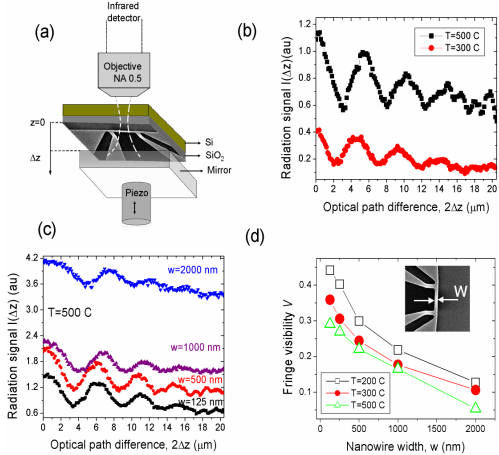

Fig 1. (a) Schematics of the experimental setup to interfere the infrared thermal radiation from a heated nanowire with its image in a movable mirror. (b) Interference fringes at 300ºC and 500º C for a 125 nm wide nanowire. (c) Interference fringes at 500º C for variable width (w) nanowires. (d) Change in the visibility of the interference fringe as function of nanowire width (w) and temperatures (T).

In analogy to Lloyd's mirror interference experiment[1] and as illustrated in Fig. 1(a), the fringes originate from the self interference of the thermal emission from the nanowire ($I_1$) and its image ($I_2$) created in the mirror. The radiation signal (I) on the detector is:

$$I = I_1 + I_2 + 2I_1I_2 \operatorname{Re}\gamma_{12}(\tau) \qquad (1)$$

where $\gamma_{12}(\tau)$ is the complex degree of coherence function and $\tau$ is the time delay between the radiation coming from the nanowire and its image. The coherence function $\gamma_{12}(\tau)$ describes the correlation in time between the nanowire radiation and the radiation from its image in the mirror and thus determines the appearance of interference fringes. The magnitude of the coherence function, $|\gamma_{12}(\tau)|$, determines the degree of coherence and its value is 1 for a perfectly coherent light source and 0 for a completely incoherent light source.[1] The thermal radiation from nanowires it is expected to be partially coherent ($|\gamma_{12}(\tau)| << 1$), with the total signal composed of interference fringes on top of an incoherent background. Typical interference traces at 300º C and 500º C for an 125 nm wide nanowire have three clearly distinguishable interference maxima over an optical path difference of 20 μm (mirror displacement of 10 μm) (Fig. 1 (b)). From the interference traces an "average" detected emission wavelength of ~5μm can be determined, using the constructive interference conditions (i.e. optical path difference or the distance between the image and the source (2Δz)) is a multiple of the wavelength, i.e. 2Δz=λ . This is consistent with Wien's displacement law, which predicts that the peak wavelength is outside the response of the InSb detector for the temperature range explored in these experiments, and thus only the high energy tail of the Planck's distribution is probed in the 3 to 5 μm wavelength region. We also note the thermal emission spectra from these nanowires can be quite different from the classical blackbody and details of the spectral distributions will be published shortly.[9]

Interference traces for wider nanowires (see Fig. 1(c)) show similar size interference signals riding on a larger incoherent background. In order to compare the interference fringes at different nanowire width, we describe the traces using the concept of the fringe visibility[1]:

$$V = \frac{2I_{\max} - I_{\min 1} - I_{\min 2}}{2I_{\max} + I_{\min 1} + I_{\min 2}}, \qquad (2)$$

where $I_{\max}$ is the value of the maximum interference peak and $I_{\min 1}$ and $I_{\min 2}$ are the two neighboring minima. This definition takes into account that the signal strength changes as the mirror moves into the depth of focus of our detection system.

The visibility of the largest maximum interference peak (at mirror sample gap of ~2.5 μm) is compared for the different size nanowires for a few selected temperatures in Fig. 1(d). Smaller and narrower nanowires show a larger fringe visibility. The fringe visibility increases for lower nanowire temperatures (longer wavelength) for a given size nanowire. For very wide nanowires, the visibility of the fringes is reduced and the radiation resembles more a blackbody radiation.

For similar ($I_1=I_2$) thermal radiation signals from the nanowire ($I_1$) and its image ($I_2$), the measured fringe visibility is equal to the degree of coherence[1]: $V = |\gamma_{12}|$. This relation allows estimating the degree of coherence using the fringe visibility results from our experiments. We note that the fringe visibility can be estimated from the interference traces at three distances, which correspond to the three visible fringes in Fig. 1(b) and Fig. 1(c).

The degree of coherence for a radiation source with a known spectral distribution $g(\nu)$ can be also calculated theoretically[10-11] by

$$\gamma_{12}(\tau) = \int_0^\infty g(\nu) e^{-2\pi i \nu \tau} d\nu \qquad (3)$$

where $\nu$ is the frequency and $\tau$ is the time delay between the interfering light sources. For a classical blackbody radiator $g(\nu)$ is given by $g(\nu) \sim \nu^3 / (\exp(\frac{h\nu}{k_B T}) - 1)$ where T is temperature and $\tau$ is the time delay determined by the optical path difference : $\tau = 2\Delta z / c$.

In Fig. 2(a), the computed degree of coherence for a classical blackbody radiator (Eq. 3) is compared with the measured degree of coherence for nanowires at 500º C as function of optical path difference. The experimental data has a large error bar in z due to the uncertainty in determining the "contact" or zero point mirror sample gap in this experiment. An estimated 4 µm gap may exist at contact point due to the tilt of the mirror and nanowire sample. It is evident from Fig. 2(a) that the thermal radiation from the nanowires is significantly more coherent than a classical blackbody radiator.

The coherence length can be determined from Fig. 2(a) as the distance where the fringe visibility (degree of coherence) reaches 10%.[12] For example, at 500º C the coherence length, $l_{BB}$, for blackbody radiation is 4 µm (Fig. 2(a)). In contrast to the classical blackbody radiator, the 125 nm wide nanowires yield a coherence length, $l_{NW}$ of ~ 20 µm (Fig. 2(a)), which is ~ 5 x longer than for the classical blackbody radiator. However, wider and larger nanowires have shorter coherence lengths, which are comparable with the blackbody radiation.

As the temperature of the nanowire increases, the degree of coherence (the coherence length) drops (Fig. 2(b)). The coherence length for the 125 nm wide nanowire is estimated to be 26 µm at 300º C and drops to 20 µm at 500º C. By way of comparison, the classical blackbody radiator has a coherence length of ~2 µm at 600 °C and ~5 µm at 300 °C.[9] We note that the actual coherence length of the nanowire thermal emission is most likely significantly longer since part of the disappearance of the interference fringes is caused simply by the fact that the mirror image of the nanowire is moved out of the depth of focus of our detection system.

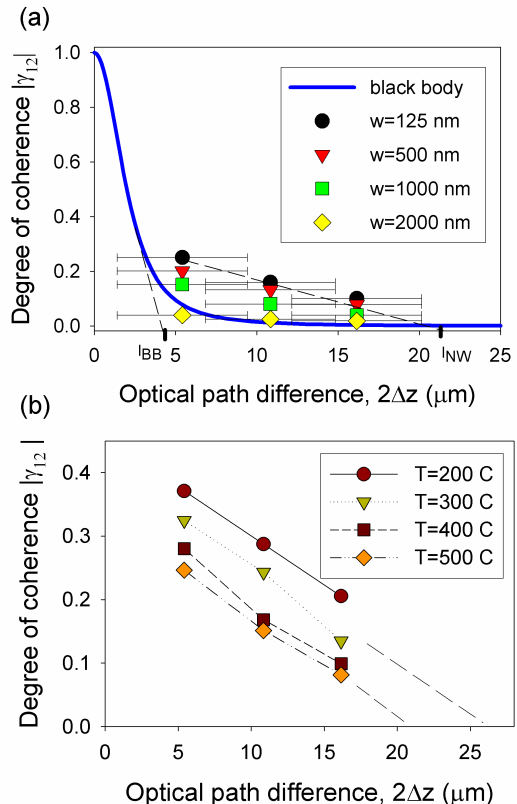

Fig 2. (a) Degree of coherence for blackbody radiation and nanowires as function of path difference between nanowire and its image. The coherence length for blackbody radiation is $l_{BB}$ ~4 µm at 500º C and $l_{NW}$ ~ 20 µm for the 125 nm wide nanowire. (b) Increase in the coherence length from 20 µm at 500º C to 26 µm at 300º C for a 125 nm wide nanowire.

In addition to increased coherence a strong increase in the radiation efficiency is observed for narrow nanowires. If the radiation signal (Fig. 1(c)) is normalized by the emitting area (10 $w^2$), the radiation intensity for narrower nanowires is highly increased compared with the wider ones (Fig. 3(a)). In general, for classic blackbody radiation emitted by large objects, the radiation intensity (radiation signal divided by emitting area) is expected to be the same for objects at similar temperatures. This is changing as the emitting object dimensions are reduced below the emission wavelength with narrower nanowires becoming enhanced thermal emitters[7]. Although this increase could be partially due to an increased spreading of heat for the smaller nanowires we note that the substrate is transparent in the 3-5 µm region of our detection system (we detect the radiation through the substrate). Taking the magnitude of radiation intensity at large gaps, where the nanowire image contribution is negligible, an almost exponential increase of the intensity for narrower nanowires is measured at different temperatures (Fig. 3(b)). The increased coherence and radiation efficiency of the thermal radiation emitted by nanowires is consistent with the picture of the confinement of charge fluctuations and increased correlation as the nanowires get narrower.[9]

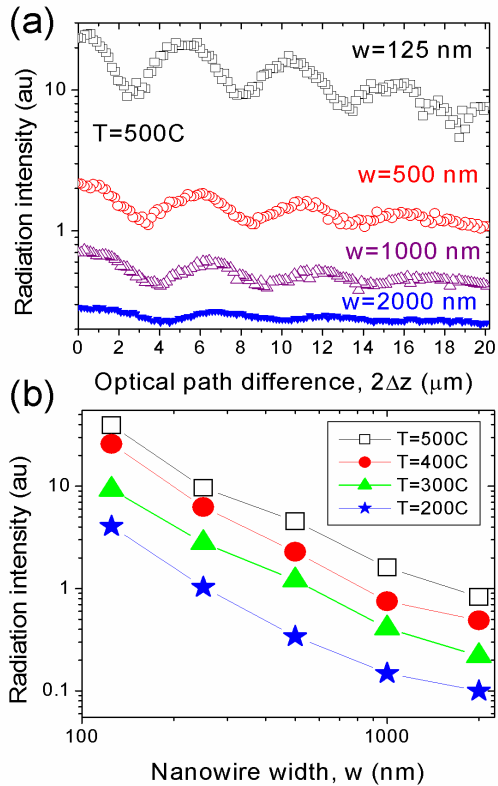

Fig 3. (a) The radiation intensity (radiation signal normalized by emitting area) is highly enhanced for narrower nanowires. (b) Enhanced thermal emission at different temperatures due to charge confinement and increase correlation in narrower nanowires.

In conclusion, the coherence properties of the thermal radiation from scaled size metallic nanowires with dimensions smaller or comparable to the emission wavelengths were investigated. A high level of coherence is observed as measured by the self-interference of the thermal radiation. Specifically, we determined a coherence length larger than 25 μm for a 125 nm wide nanowire at 300º C, much larger than the coherence length of the blackbody radiation. The coherence of the thermal radiation increases as the width of nanowire and its temperature is decreased. In addition to increased coherence, a more efficient thermal radiation emission is observed for narrower nanowires.